\documentclass[structabstract]{aa}
\usepackage{graphicx}
\usepackage{url}
\usepackage{natbib}
\usepackage{lineno}

\title{A stacking method to study the gamma-ray emission of source samples based on the co-adding of Fermi LAT count maps}
\authorrunning{B. Huber et al.}
\titlerunning{A stacking method to study the gamma-ray emission of source samples}

\author{B.~Huber\inst{\ref{inst1},\ref{inst2}}
\and C.~Farnier\inst{\ref{inst2},\ref{inst3}}
\and A.~Manalaysay\inst{\ref{inst1}}
\and U.~Straumann\inst{\ref{inst1}}
\and R.~Walter\inst{\ref{inst2}}}

\institute{Physik - Institut, Universit\"at Z\"urich, Winterthurerstrasse 190, CH-8057 Z\"urich, Switzerland\label{inst1}\\
\email{ben.huber@physik.uzh.ch}
\and ISDC Data Center for Astrophysics, Center for Astroparticle Physics, Observatory of Geneva, University of Geneva, Chemin d'Ecogia 16, CH-1290 Versoix, Switzerland\label{inst2}
\and Oskar Klein Center, Department of Physics, Stockholm University Center, Albanova University Center,\\ SE-10691 Stockholm, Sweden\label{inst3}}

\date{submitted to Astronomy \& Astrophysics}

\abstract {} {We present a stacking method that makes use of co-added maps of gamma-ray counts 
produced from data taken with the Fermi Large Area Telescope. 
Sources with low integrated gamma-ray fluxes that 
are not detected individually may become detectable when their corresponding 
count maps are added.} {The combined data set is analyzed with a maximum likelihood method
taking into account the contribution from point-like and diffuse background sources. 
For both simulated and real data, detection significance and integrated gamma-ray flux 
are investigated for different numbers of stacked sources using the public Fermi \textit{ScienceTools} for analysis and data preparation.} {The 
co-adding is done such that potential source signals add constructively, in contrast to the signals from background sources, 
which allows the stacked data to be described with simply structured models. We show, for different scenarios, that the stacking method 
can be used to increase the cumulative significance of a sample of sources and to characterize the corresponding gamma-ray emission.
The method can, for instance, help to search for gamma-ray emission from galaxy clusters.} {}

\keywords{Methods: data analysis, statistical, Gamma-rays: general}

\begin{document}
\maketitle

\section{Introduction}
\label{introduction}
In conjunction with the Energetic Gamma-ray Experiment Telescope (EGRET) onboard the 
Compton Gamma-ray Observatory, 
several studies made use of a stacking method that is based on the adding (co-adding, stacking) of maps 
of gamma-ray counts and on 
a subsequent analysis with a maximum likelihood method.
These studies were performed to search 
for gamma-ray emission from, for instance, clusters of galaxies \citep{reimer03}, radio and Seyfert galaxies \citep{cillis04}, 
infrared galaxies \citep{cillis05} and potential gamma-ray sources at low galactic latitudes \citep{cillis07}.\\
\indent The basic idea of the co-adding is to add up the data such that 
the sources of interest are spatially correlated with one other, in contrast to the background 
sources within the source region. Co-adding can thus increase the signal-to-background 
ratio, resulting in an increased cumulative significance of the sources in the sample.\\
\indent Inspired by the EGRET stacking effort, we present a 
co-adding method for data obtained with the Fermi Large Area Telescope (LAT).
The LAT is a pair-conversion telescope 
onboard the Fermi Gamma-ray Space Telescope \citep{atwood09} that is capable to detect gamma rays 
with energies from 20 MeV to more 
than 300 GeV. In comparison to 
EGRET, the LAT achieves a point source sensitivity that is increased by two orders of 
magnitude \citep{egretlat}.\\
\indent The public Fermi \textit{ScienceTools} provide the stacking tool \textit{Composite2} 
that has recently been applied, for instance, to Milky way satellite galaxies to search for signals 
from dark matter annihilations \citep{garde11}. 
In contrast to the \textit{Composite2} method that makes use of summed $\log$-likelihood functions,  
the co-adding method presented in this paper is based on added maps of counts instead. 
This may increase the signal-to-background ratio for the sources of interest and 
a visible excess of counts may be achieved for the stacked signal.\\
\indent This paper is structured as follows.
In section \ref{coadding}, the co-addition of count maps is described 
together with the resulting likelihood analysis. 
In section \ref{simulation}, the method is tested with simulated data. 
As a first step, the stacking of diffuse background is studied, and significance 
and integrated photon flux upper limits of a hypothetical central point-like source 
are computed for different numbers of stacked sources. 
In a second step, the ability to detect point-like emissions with low integral photon fluxes is tested. 
For this purpose, a simulated faint point-like source is 
added at the center of each region and the stacking is repeated, investigating the development of source significance and 
integrated photon flux as well as the dependence on the spectral shape of the source emission. 
In section \ref{realdata}, the method is applied to real data, first to regions that are free of detected central point-like emission 
and second to regions that host a known point-like source at their center, that is present 
in the LAT 2-year point source catalog \citep{abdo11}.
In section \ref{discussion}, we discuss the co-adding method and its performance.\\
\indent For the analysis and data preparation in the following sections, we make use of 
the \textit{ScienceTools} (v9r23p1) \citep{scitools}, 
together with the \textit{P7SOURCE\_V6} LAT response functions.\\
\indent The scripts used for the co-adding analyses in the following sections will become 
publicly available in the future, or can be directly obtained by the corresponding author.

\section{The co-adding method}
\label{coadding} 
The co-adding method presented in this paper is based on analyzing stacked data histograms 
with a maximum likelihood method.  The contents of each individual histogram include contributions from two classes 
of background sources and potentially one signal source. 
The reason for separating the background contributions into two classes will become clear in section \ref{coaddingdata}.
Let $n_{im}$ denote the content of bin $i$ in histogram $m$, 
and $b_{im}^{(1)}$ be the estimated contribution from the first class of backgrounds in the same bin.
The first class of background is subtracted from the original histograms individually, 
before the stacking. The bin values $\widetilde{n}_{i}$ of the co-added histogram are then given by:
\begin{eqnarray}
\widetilde{n}_{i} = \sum_{m=1}^{j}(n_{im}-b_{im}^{(1)}) {\rm{\,,}}
\end{eqnarray}
where $j$ is the number of stacked histograms.
These bin values are fit by a model $\widetilde{\theta}_{i}$ which is the set of bin contents that include a prediction for the 
potential signal source, $s_{im}$, combined with a prediction for the second class of background events, $b_{im}^{(2)}$:
\begin{eqnarray}
\widetilde{\theta}_{i} = \sum_{m=1}^{j}(s_{im}+b_{im}^{(2)}) {\rm{\,.}}
\end{eqnarray}
The normalization of $s_{im}$ and $b_{im}^{(2)}$ are considered to be free parameters during the fit 
and varied until a maximum value for the standard Poisson log-likelihood \citep{mattox96},
\begin{eqnarray}
\label{logLike_coadded}
	\log\widetilde{\mathcal{L}} = \sum_{i}\left(\widetilde{n}_{i}\cdot \log\widetilde{\theta}_{i}\right) - \sum_{i}\widetilde{\theta}_{i} {\rm{\,,}}
\end{eqnarray}
is found, giving the best-fit value for the amplitude of the potential signal. 
For each $j\in[1,j_{\rm{max}}]$, where $j_{\mathrm{max}}$ is the total number of available histograms, the signal amplitude and the 
corresponding statistical significance are computed. True detections are expected to show an increasing trend in significance as more 
histograms are added to the stack, 
while false positives should result in cumulative significances that remain low.\\
\indent In the following, each histogram represents a region of interest (ROI) and is given by a 
three-dimensional histogram called a CountCube.\\
The first class of background sources, $b_{im}^{(1)}$, corresponds to known gamma-ray point sources, as they are listed 
in the LAT 2-year point source catalog \citep{abdo11},
while the second class, $b_{im}^{(2)}$, refers to diffuse galactic and extragalactic gamma-ray emissions.

\subsection{The likelihood function of the standard Fermi analysis}
\label{standardana}
When the Fermi spacecraft is operated in survey mode, the LAT obtains full sky coverage. 
Photons that belong to a ROI on the sky, are selected with the tool \textit{gtselect} 
which performs the basic regional cuts as well as the selection of defined intervals for observation
time and energy (for more details, see sections \ref{simulation} and \ref{realdata}).
Here, the selection is done such that the sources of interest are positioned at the center of the corresponding ROIs. 
The ROI size is chosen large enough to account for the point spread function (PSF) of the instrument 
\citep[see e.g.][for details on the PSF]{atwood09}. 
Additional time cuts are performed with the tool \textit{gtmktime} which takes the pointing and position history
of the spacecraft into account and makes sure that only good time intervals are used for the analysis. This removes, 
for instance, events taken when the spacecraft passes through the South Atlantic Anomaly and
photons that come from the earth's limb \citep{petry05}.
Afterwards, the tool \textit{gtbin} is applied to fill 
the photons into a CountCube. Two dimensions in the CountCube are
for the sky position, e.g. right ascension and declination, whereas the third dimension corresponds 
to the reconstructed energy.\\
\indent The $\log$-likelihood function that is used for the standard binned analysis of single (not stacked) ROIs
is given by \citep{mattox96}:
\begin{eqnarray}
\label{loglike}
	\log \mathcal{L} = \sum_{i}\left(n_{i}\cdot \log\theta_{i}\right) - \sum_{i}\theta_{i}
\end{eqnarray}
where $n_{i}$ are the measured number of photon counts for bin $i$ as they are stored in the CountCube,  
and $\theta_{i}$ are the number of counts predicted by a model for the same bin.
The index $i$ runs over all bins in the CountCube.\\
\indent The model that is used to predict the $\theta_{i}$ contains the positions and spectral shapes 
of known point sources in the ROI, and it includes the expected 
isotropic contribution from the extragalactic diffuse background (EGB) 
and the region-dependent
contribution from the galactic diffuse background (GB).
This model is converted into the model-predicted number of counts with the aid of a SourceMap that takes 
the integrated exposure time during the observation and the instrument response functions, 
mainly PSF, effective area and energy-dependent corrections, 
into account and provides the appropriate conversion factors for each bin $i$.
For every source in the model an individual SourceMap is generated by the tool \textit{gtsrcmaps}, 
using the same spatial and energy binning as in the underlying CountCube.
The model-predicted number of counts $\theta_{i}$ in bin $i$ are calculated through:
\begin{eqnarray}
\label{predictedcounts}
	\theta_{i} = N_{\rm{GB}}\cdot F_{{\rm{GB}},i}\cdot S_{{\rm{GB}},i} + N_{\rm{EGB}}\cdot F_{{\rm{EGB}},i}\cdot S_{{\rm{EGB}},i} \\ \nonumber
	+ \sum_k \left[N_{0,k}\cdot F_{k,i}(\alpha,...)\cdot S_{k,i}\right]{\rm{\,,}}
\end{eqnarray}
where $S_{{\rm{GB}},i}$, $S_{{\rm{EGB}},i}$ and $S_{k,i}$ represent the SourceMap values for 
the GB, EGB and the point sources $k$, that belong to the sky position and the energy interval of bin $i$. 
$N_{\rm{GB}}\cdot F_{{\rm{GB}},i}$ and $N_{\rm{EGB}}\cdot F_{{\rm{EGB}},i}$ denote the photon fluxes predicted for bin $i$, 
where $N_{\rm{GB}}$ and $N_{\rm{EGB}}$ are the corresponding normalization parameters that are free during the likelihood fit.
While $F_{{\rm{EGB}},i}$ is uniform for all the bins of a given energy and is derived from a 
fixed spectrum \citep{EGBmodel}, $F_{{\rm{GB}},i}$
is derived from a three-dimensional distribution map of differential photon fluxes \citep{GBmodel}.
For this reason, the SourceMap of the galactic diffuse emission, that is created with the \textit{ScienceTools}, incorporates the factor 
$F_{{\rm{GB}},i}$. In the following, $S^{\rm{impl}}_{{\rm{GB}},i} = F_{{\rm{GB}},i}\cdot S_{{\rm{GB}},i}$
denotes the SourceMap of the GB model as it is implemented in the \textit{ScienceTools}.
The photon flux of source $k$ is denoted by $N_{0,k} \cdot F_{k,i}(\alpha,...)$, based on 
a source spectrum that may depend on several 
parameters, e.g. a prefactor $N_0$ and a photon index $\alpha$ in the case of a power-law spectrum \citep[see][]{sourcemodel}.

\subsection{Co-adding of data}
\label{coaddingdata}
ROIs, selected and prepared as previously described, are combined by adding up
their corresponding CountCubes.
For this purpose, we introduce a new coordinate system in which the origin is defined to be at the center of
the combined ROI. All sources of interest are thus located at the origin.
The goal is to analyze the co-added data with the maximum likelihood method implemented in the \textit{ScienceTools}.
We model the co-added diffuse backgrounds, GB and EGB, by building a weighted sum of the SourceMaps.
This takes into account that the contributions of the diffuse backgrounds are different for each ROI and 
that the co-adding adds up the exposures.
$\widetilde{S}_{{\rm{GB}},i}$ and $\widetilde{S}_{{\rm{EGB}},i}$ are the stacked SourceMaps for the GB and EGB model, respectively:
\begin{eqnarray}
	\label{sourcemap_diffuseBG_coadded}
	\widetilde{S}_{{\rm{GB}},i} &=  \sum_{m} \left[N_{\rm{GB}}\cdot S^{\rm{impl}}_{{\rm{GB}},i}\right]_m {\rm{\,and}}\\
	\label{sourcemap_diffuseBG_coadded2}
	\widetilde{S}_{{\rm{EGB}},i} &=  \sum_{m} \left[N_{\rm{EGB}}\cdot S_{{\rm{EGB}},i}\right]_m{\rm{\,,}}	
\end{eqnarray}
where $m$ denotes the different ROIs. The factors $N_{\rm{GB}}$ and $N_{\rm{EGB}}$ are used to normalize the SourceMaps 
according to the diffuse background contributions in each region, known from individual region analyses (see below).\\
\indent Depending on the ROI size, the number of detected point sources increases rapidly
and leads for the co-adding to a model with a large number of components. 
In order to keep the model simply structured, we follow a different strategy.
Before co-adding the data, a binned likelihood analysis \citep{blikelihood} is performed on each individual ROI using 
models that contain the diffuse backgrounds and the Fermi-LAT detected point sources from the LAT 2-year point source catalog, 
within and close to the ROIs.   
These analyses yield the normalizations of the diffuse backgrounds and the parameters of the point sources for 
each individual ROI. More than 99\% of the sources in the LAT 2-year point source catalog are characterized by a point-like 
gamma-ray emission, only very few sources may appear extended with extensions on the order of the PSF. 
In the following, we treat all point sources that are included in the models as point-like emissions.
All Fermi-LAT detected point sources are then declared as background and simulated using the tool \textit{gtobssim}, 
in order to subtract them from 
the measured data. This leads to a simpler form of equation \ref{predictedcounts}, in which the last term is suppressed. 
The resulting CountCubes, that is the data minus the simulated point sources, are finally co-added.\\ 
\indent In order to investigate sources at the center 
of the ROIs, the model that we apply for the analysis of the co-added CountCubes contains, besides the diffuse backgrounds, 
a common source of interest at the ROI center, that is denoted as test source.
For the analyses in the following sections, the test source is described as a point-like source located at the ROI center
using a power-law spectrum with a prefactor $\widetilde{N}_0$ and a photon index $\widetilde{\alpha}$ as model parameters \citep[see][]{sourcemodel}.
Similar to the case of the diffuse backgrounds, the SourceMaps for the test source are first produced individually for 
each ROI $m$ and then
added up in order to take the total exposure into account. The bin values of the co-added SourceMaps for the 
test source, $\widetilde{S}_{\rm{test\,source},i}$,
are then given by:
\begin{eqnarray}
	\label{sourcemap_testsource_coadded}
	\widetilde{S}_{{\rm{test\,source}},i} =  \sum_m \left[S_{{\rm{test\,source}},i}\right]_m{\rm{\,.}}
\end{eqnarray}
\indent From the stacked SourceMaps defined in equations \ref{sourcemap_diffuseBG_coadded}, \ref{sourcemap_diffuseBG_coadded2} and 
\ref{sourcemap_testsource_coadded}, the number of co-added model-predicted counts $\widetilde{\theta}_{i}$ is derived by:
\begin{eqnarray}
	\label{predictedcounts_coadded}
	\widetilde{\theta}_{i} = \widetilde{N}_{\rm{GB}} \cdot \widetilde{S}_{{\rm{GB}},i} + \widetilde{N}_{\rm{EGB}}\cdot F_{{\rm{EGB}},i}\cdot \widetilde{S}_{\rm{EGB},i} \\ \nonumber
	+ \widetilde{N}_0 \cdot \widetilde{F}_{{\rm{test\,source}},i}(\widetilde{\alpha})\cdot \widetilde{S}_{{\rm{test\,source}},i}\rm{\,,}
\end{eqnarray}
where the normalizations of the diffuse backgrounds $\widetilde{N}_{\rm{GB}}$, $\widetilde{N}_{\rm{EGB}}$ and 
the test source parameters, prefactor $\widetilde{N}_0$ and photon index $\widetilde{\alpha}$, 
may be free during the likelihood fit. 
Due to the normalized sum of SourceMaps in equations \ref{sourcemap_diffuseBG_coadded} and \ref{sourcemap_diffuseBG_coadded2}, 
the values of $\widetilde{N}_{\rm{GB}}$ and $\widetilde{N}_{\rm{EGB}}$ are expected to be close to 1 and $F_{{\rm{EGB}},i}$
is derived from the same fixed spectrum \citep{EGBmodel} as in equation \ref{predictedcounts}.
Since the SourceMaps of the test source are summed in equation \ref{sourcemap_testsource_coadded}, the test source parameters
and the resulting flux $\widetilde{N}_0\cdot \widetilde{F}_{{\rm{test\,source}},i}$ represent values that are averaged 
by the total stacked exposure.\\ 
\indent The $\log$-likelihood function of the co-adding analysis $\log\widetilde{\mathcal{L}}$ is given 
in equation \ref{logLike_coadded}, where $\widetilde{n}_{i}$ is the number of co-added measured counts in bin $i$ after the 
subtraction of the simulated 
point-like sources. 
Both equations \ref{loglike} and \ref{logLike_coadded} represent a likelihood that is derived from Poisson-distributed
numbers of observed counts. The subtraction of the simulated counts transforms, for a fraction of bins, 
the Poisson distribution into a Skellam probability distribution 
\citep{Skellam46}. For regions located at galactic latitutes $|b|>25^\circ$ (to avoid the high number of 
contributing background sources close to the galactic plane and to be consistent with the analyses performed in sections \ref{simulation} 
and \ref{realdata}), about $3\%$ of the bins in a CountCube are affected by the subtraction. Therefore, 
the likelihood defined in equation \ref{logLike_coadded} is used for the co-adding analysis in good approximation.\\
\indent We perform the binned likelihood analysis of the co-added data through the likelihood python interface of the 
\textit{ScienceTools} \citep{pylike} using the minimizer \textit{Minuit} \citep{minuit}.
During the likelihood analysis, equation \ref{logLike_coadded} is maximized, which 
results in maximum likelihood estimators for the free parameters in the applied model. 
As a measure of the test source significance, we compute the test statistic 
${\rm{TS}} = -2(\log \mathcal{L}_0 - \log \mathcal{L}_1)$, in which $\mathcal{L}_0$ and $\mathcal{L}_1$ are the 
maximized likelihood-values given that only the diffuse backgrounds are present in the model (null hypothesis) and that 
a test source is present in addition to the diffuse backgrounds (alternative hypothesis), respectively.\\
\indent For the analyses in the following sections, the prefactor $\widetilde{N}_0$ is the only free parameter 
in the spectral model of the test source.
Hence, if the photons were only due to the background fluctuations from the sources defined in the null hypothesis,
then the TS values would follow approximately a $\chi^2_k/2$-distribution with $k=1$ free parameter. 
The significance level of the test source can then be denoted as $\sqrt{{\rm{TS}}}$ in units of sigma ($\sigma$) 
\citep{mattox96,pdg10}, based on a one-sided Gaussian quantile.
Simulations showed that a $\chi^2_1/2$-distribution is still true
when simulated point sources are subtracted from the data and when stacking is performed.
In the following sections, we use a detection threshold of TS$\ge$25, corresponding to a 
5$\sigma$ detection level, based on one 
more free parameter in the alternative hypothesis compared to the null hypothesis.

\section{Tests with simulated data}
\label{simulation}
In the following, the co-adding is tested using 40 simulated ROIs that are randomly distributed at 
high galactic latitudes $|b|>25^\circ$. 
Sources close to the galactic plane are excluded to avoid potential mis-modelings of the galactic diffuse 
emissions and to avoid a high number of background point sources in the analyzed ROIs.
This is mainly relevant for the analysis of real data in section \ref{realdata} and applied here for consistency. 
The simulations are performed with $gtobssim$ for 162 weeks of Fermi operation 
using the real spacecraft information (2008-08-04 to 2011-09-13) and for energies from 200 MeV to 100 GeV.
The resulting photon data is filled into CountCubes with 100 x 100 pixels and 40 energy intervals in logarithmic scale, 
corresponding to square-shaped ROIs with an angular sidelength of 20$^\circ$.\\ 
\indent We made use of 40 ROIs because the available number of real sources used to test this method in section 
\ref{realsources} is on the order of 30. 
Furthermore, the co-adding method will be applied 
to search for gamma-ray emission from a sample of galaxy clusters in an upcoming paper (Huber et al. 2012, in preparation). 
The available number of clusters, after selection cuts similar to those used by \citep{reimer03}, is expected to be on the same order.

\subsection{Robustness against false detections}
\label{falsedetections}
\begin{figure}
\resizebox{\hsize}{!}{\includegraphics[width=\linewidth]{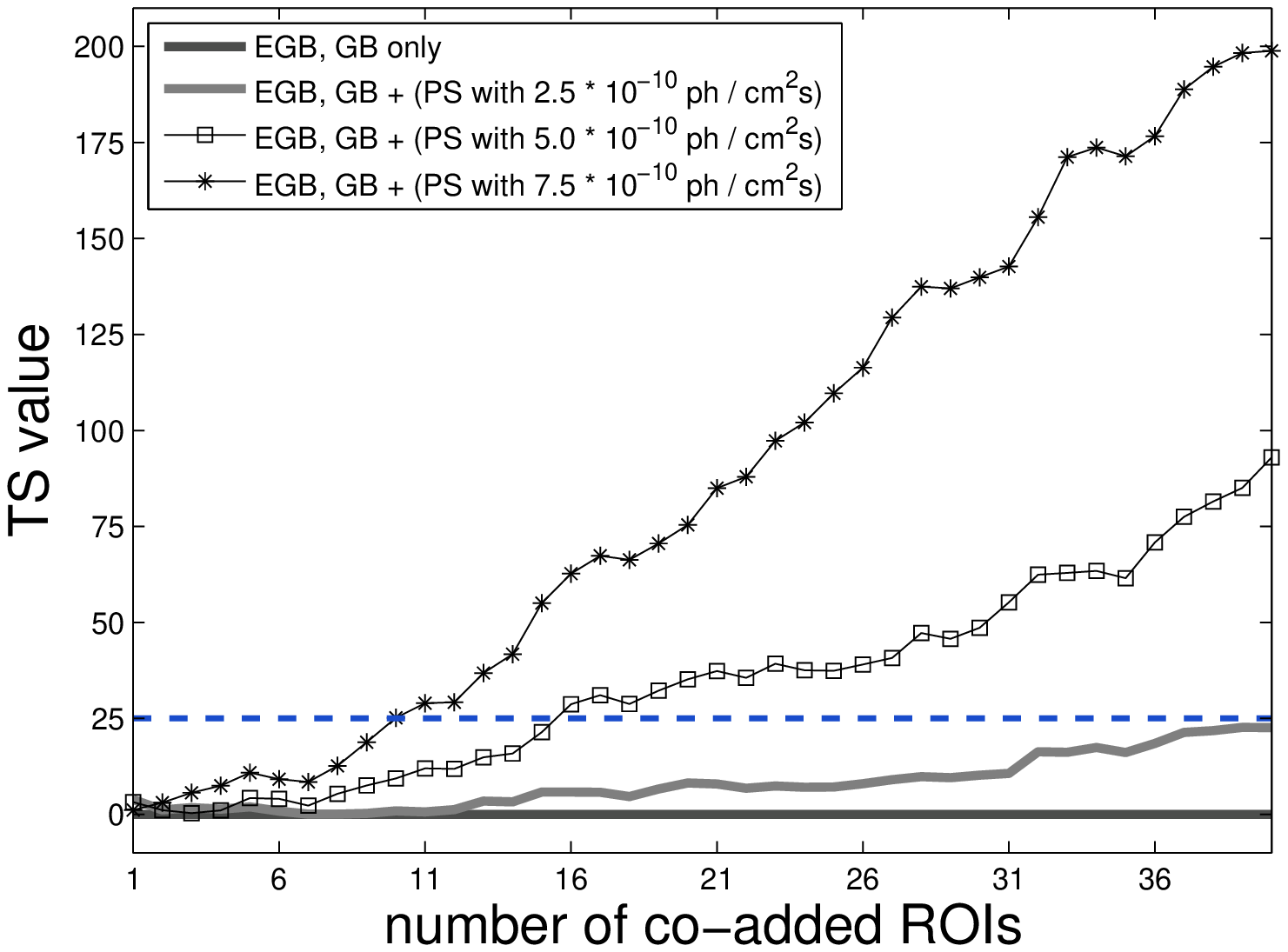}}
\caption{The test statistic values with respect to point-like emission at the center of the ROIs versus the number of co-added ROIs.
		The co-adding is performed for simulated regions that only contain EGB and GB and 
		for regions that additionally contain a simulated point-like source (PS) at the center. 
		Three samples with different integrated fluxes of the point-like source, $2.5 \times 10^{-10}$,
		$5.0\times 10^{-10}$ and $7.5\times 10^{-10}$ ph/(cm$^2$s), are used.
		The detection threshold TS $\ge$ $25$ is indicated by the dashed line.}
\label{SimROI_TS}
\end{figure}

First, we investigate the probability of a false detection of a point-like emission
due to statistical fluctuations of the co-added diffuse emissions. 
The model used to describe the test source at the ROI center is defined by a power law 
with a fixed photon index of $-2.0$. The normalizations of EGB and GB
and the prefactor of the test source are free parameters during the likelihood fit.
As expected, co-adding of diffuse background yields no significant signal at the position of the test source.
As shown by the dark grey solid line in figure \ref{SimROI_TS}, the TS values take values close to zero
for any number of co-added ROIs. Although this curve appears flat, the TS values underlie the fluctuations
expected for the analysis of pure diffuse background.
Since the TS values stay below TS=$25$, we compute the 90\% confidence level (CL) upper limits (UL). 
To obtain the 
90\% CL UL on the gamma-ray flux, the prefactor of the test source spectrum is stepwise-increased until 
$(\log \mathcal{L}_{\rm{max}} - \log \mathcal{L}_{\rm{max}}|_{\widetilde{N}_0^{\rm{inc}}}) =2.71/2$ \citep{cowan},
where $\mathcal{L}_{\rm{max}}$ is the maximized likelihood-function and 
$\mathcal{L}_{\rm{max}}|_{\widetilde{N}_0^{\rm{inc}}}$ is the 
likelihood-function recomputed after $\widetilde{N}_0$ has been incremented. The resulting test source spectrum 
is then integrated over the full simulated energy range from 200 MeV to 100 GeV.\\ 
\indent The UL development for the test source is shown by the dark grey solid line in figure \ref{SimROI_flux}. 
With an increasing number of co-added ROIs the probability of false signals due to Poissonian fluctuations is reduced. 
The computed UL decreases until it reaches an asymptote after approximately 25 co-adding steps. 
By co-adding 40 ROIs, an upper limit on the integrated flux of approximately $3\times 10^{-11}$ ph/(cm$^2$s) is obtained.

\subsection{Detectability of weak signals}
\label{weaksignals}
\begin{figure}
\resizebox{\hsize}{!}{\includegraphics[width=\linewidth]{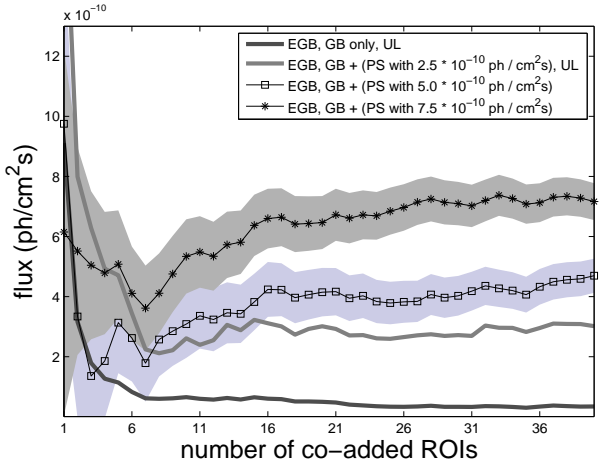}}
\caption{The integrated photon flux or 90\% CL upper limit (UL) with respect to point-like emission at the center of the ROIs versus the number of co-added ROIs.
		The co-adding is performed for simulated regions that only contain EGB and GB and 
		for regions that additionally contain a point-like source (PS) at the center. 
		Three samples with different integrated fluxes of the point-like source, $2.5 \times 10^{-10}$,
		$5.0\times 10^{-10}$ and $7.5\times 10^{-10}$ ph/(cm$^2$s), are used. 
		The grey and blue shaded areas correspond to the statistical uncertainties
		on the integrated flux values.}
\label{SimROI_flux}
\end{figure}

In a second step, we test the ability of the method to detect a signal from weak point-like emissions.
Additional sets of simulations are performed that add a point-like source, denoted as central source,
to the center of the previously simulated diffuse background regions. The central sources are simulated 
for different integrated fluxes [$2.5$, $5.0$ and $7.5 \times 10^{-10}$ ph/(cm$^2$s)] using a power-law 
spectrum with a photon index $-2.0$.
Standard analyses of the individual ROIs (no co-addition applied) reveal no significant signal (TS $\ge$ $25$) associated 
with the test source for any of these data sets. 
We perform the co-adding method independently 
for the different integrated fluxes using a power-law test source with a photon index $-2.0$, 
for which the resulting developments of the TS values are shown in figure \ref{SimROI_TS}.
Central sources with a flux of $2.5 \times 10^{-10}$ ph/(cm$^2$s) remain undetected during 40 co-adding steps.
In contrast, sources with twice this integrated flux yield a clear detection after 15 co-added ROIs, and 
sources with a flux of $7.5 \times 10^{-10}$ ph/(cm$^2$s) are detected after 10 co-additions. 
The high TS values, on the order of 100 and 200, after 40 stacking steps show clearly the power of this method
to detect weak emissions from combined regions.\\
\indent In each co-adding step, we also compute integrated flux values for the detected and 90\% CL UL for the undetected 
samples, which are shown in figure \ref{SimROI_flux}.
The UL values for the sources with $2.5 \times 10^{-10}$ ph/(cm$^2$s) decrease strongly at the beginning and reach an asymptote 
after 10 co-adding steps, that is significantly higher than the one obtained from the diffuse background. 
This can be explained
by the fact that the co-added central sources in this sample almost reach detection level.
It can be seen from the same figure that the statistical uncertainties (grey and blue shaded areas) on the
integrated fluxes of the detected samples are reduced with an increasing number of co-additions.

\subsection{Dependence on the spectral hardness}
\begin{figure}
\resizebox{\hsize}{!}{\includegraphics[width=\linewidth]{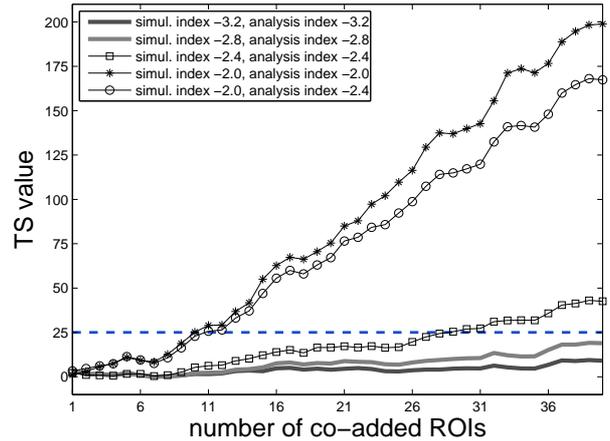}}
\caption{The test statistic values with respect to point-like emission at the center of the ROIs versus the number of co-added ROIs.
		The stacking is performed for simulated ROI samples 
		that contain central sources with photon indices $-2.0$, $-2.4$, $-2.8$ and $-3.2$.
		For the analysis, power-law spectra are used that apply the same fixed photon indices as in the 
		corresponding simulations.
		The stacking is also performed for ROIs that contain central sources with photon index  
		$-2.0$ using a test source with photon index $-2.4$ in the analysis.}
\label{SimROI_4Indices_TS}
\end{figure}

\begin{figure}
\resizebox{\hsize}{!}{\includegraphics[width=\linewidth]{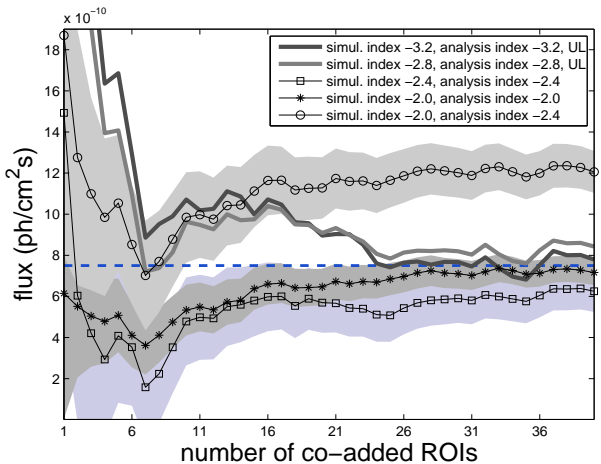}}
\caption{The integrated photon flux or 90\% CL upper limit (UL) with respect to point-like emission at the center of the ROIs versus the number of co-added ROIs.
		The stacking is performed for simulated ROI samples 
		that contain central sources with photon indices $-2.0$, $-2.4$, $-2.8$ and $-3.2$.
		For the analysis, power-law spectra are used that apply the same fixed photon indices as in the 
		corresponding simulations.
		The stacking is also performed for ROIs that contain central sources with photon index  
		$-2.0$ using a test source with photon index $-2.4$ in the analysis.
		The grey and blue shaded areas correspond to the statistical uncertainties on the integrated flux values.
		All central sources are simulated with an integrated flux 
		of $7.5\times 10^{-10}$ ph/(cm$^2$s), indicated by the dashed line.}
\label{SimROI_4Indices_flux}
\end{figure}

In the following, the dependence of the source significance on different spectral shapes is tested.
For this, four samples of simulated ROIs in which the central sources have a common integrated flux of 
$7.5\times 10^{-10}$ ph/(cm$^2$s) but different photon indices [$-2.0$, $-2.4$, $-2.8$ and $-3.2$] were produced.
The samples are co-added and analyzed separately using a test source model that applies the same spectral shape
as in the respective simulation. The resulting TS values are shown in figure \ref{SimROI_4Indices_TS}.
The two samples with photon indices $-2.0$ and $-2.4$ are clearly detected after 10 and 28 co-additions, respectively.
Although the simulated integrated flux is the identical, we find no significant signal for the two softer spectra. 
Hard spectra provide, compared to soft spectra, an increased number of events in high energy bins, which make the likelihood, 
due to an improved PSF at high energies, more sensitive to a spatial correspondence between model and data.\\ 
\indent In figure \ref{SimROI_4Indices_flux}, the integrated flux of the test source is reported for the samples with photon indices $-2.0$ 
and $-2.4$, for which the values obtained after 40 co-additions are consistent with the simulated value of $7.5\times 10^{-10}$ ph/(cm$^2$s). 
Since a TS $\ge$ $25$ is not reached for the samples with photon indices $-2.8$ and $-3.2$, 90\% CL flux UL are computed instead.\\
\indent In cases in which the exact spectral features in the data are unknown, it may help to investigate
the data using different spectral shapes.
In figure \ref{SimROI_4Indices_TS}, the TS values are computed for sources
simulated with a photon index of $-2.0$ and analyzed with a photon index of $-2.4$.
Applying an analysis spectrum that is softer than the spectrum in the data decreases the 
TS increment per stacking step, which is also the case if a harder analysis spectrum, e.g  with a photon index of $-1.6$, is used.
The method is thus sensitive to the correspondence between the spectral shape in the model and the spectral shape 
in the data, which is a useful feature to investigate the combined spectrum of the sources in the sample.\\ 
\indent The sensitivity of the likelihood analysis to events in high energy bins leads in this case to
an increased integrated test source flux if a photon index $-2.4$ is used, as it is shown in figure \ref{SimROI_4Indices_flux}, 
and to a decreased flux applying a photon index of $-1.6$.

\subsection{Detectability of a source sample with mixed spectral shapes}
\begin{figure}
\resizebox{\hsize}{!}{\includegraphics[width=\linewidth]{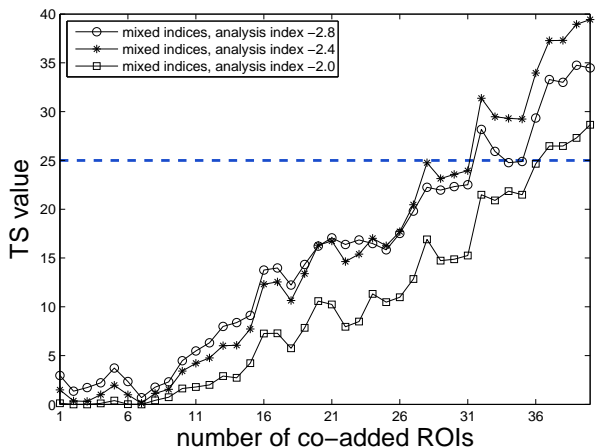}}
\caption{The test statistic values with respect to point-like emission at the center of the ROIs versus the number of co-added ROIs.
		The stacking is performed for a simulated source sample with mixed spectral shapes 
		using analysis test sources with photon indices $-2.0$, $-2.4$ and $-2.8$.
		The detection threshold TS $\ge$ $25$ is indicated by the dashed line.}
\label{SimROI_4Indices_mixed_TS}
\end{figure}

\begin{figure}
\resizebox{\hsize}{!}{\includegraphics[width=\linewidth]{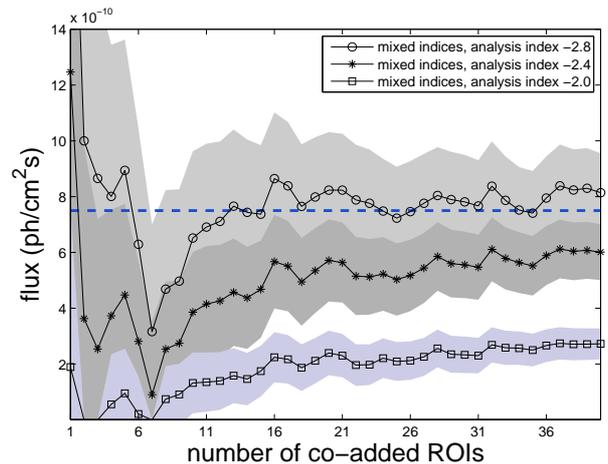}}
\caption{The integrated photon flux with respect to point-like emission at the center of the ROIs versus the number of co-added ROIs.
		The stacking is performed for a simulated source sample with mixed spectral shapes 
		using analysis test sources with photon indices $-2.0$, $-2.4$ and $-2.8$.
		The grey and blue shaded areas correspond to the statistical uncertainties on the integrated flux values.
		All central sources are simulated with an integrated flux 
		of $7.5\times 10^{-10}$ ph/(cm$^2$s), indicated by the dashed line.}
\label{SimROI_4Indices_mixed_flux}
\end{figure}

In the following, we investigate the impact of having a source sample with mixed spectra,
as it might be the case for different astrophysical sources.
For this, simulated source spectra with different photon indices [$-2.0$, $-2.4$, $-2.8$ and $-3.2$] are 
mixed during the co-adding.
We analyze the mixed sample for three different photon indices of the test source, [$-2.0$, $-2.4$, $-2.8$],
and compute the TS values after each stacking step. The common 
integrated flux of the simulated central sources is again $7.5\times 10^{-10}$ ph/(cm$^2$s). 
The results are shown in figure \ref{SimROI_4Indices_mixed_TS}.
All three cases reach the detection threshold after 30 to 35 stacking steps.
The co-adding hence 
allows sources to be detected even if different spectral shapes are involved.
The analysis with photon index $-2.4$ yields the highest significance, 
indicating that this spectral shape yields, in this example, the best agreement with the mixed spectrum. 
The corresponding integrated test source fluxes for each stacking step are shown in figure \ref{SimROI_4Indices_mixed_flux}.
The flux that we find analyzing the mixed source spectrum with a photon index of $-2.4$
is slightly below the input flux of $7.5\times 10^{-10}$ ph/(cm$^2$s).
Relative to this value, we obtain an increased and decreased integrated flux using photon indices $-2.8$ and $-2.0$, respectively, 
which can be explained again by the sensitivity of the likelihood fit to events in high energy bins as discussed previously.\\
\indent It is possible to detect the spectral shape in the data using the photon index as an 
additional free parameter during the fitting procedure.
This works well if sources with the same spectral shapes are stacked. In case of the present mixed spectra, however, it is not possible 
to achieve a sufficient likelihood fit quality if both the prefactor and the spectral index are free to vary. 

\section{Tests with real data}
\label{realdata}
\subsection{Robustness against false detections}
\begin{figure}
\resizebox{\hsize}{!}{\includegraphics[width=\linewidth]{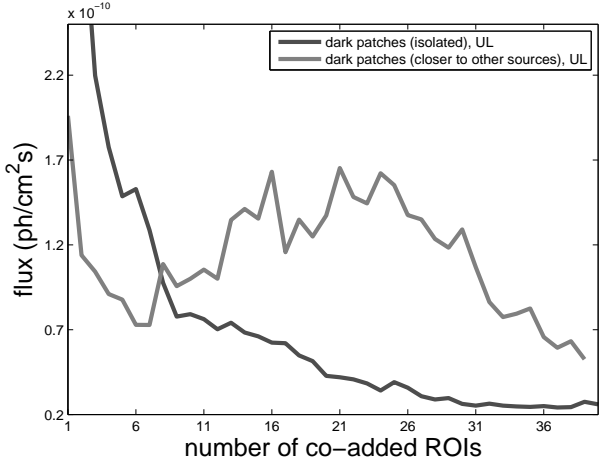}}
\caption{The 90\% CL upper limit (UL) on the integrated photon flux with respect to point-like emission at the center of the ROIs versus the number of co-added ROIs.
		The first sample consists of isolated dark patches that contain as few detected Fermi sources as possible.
		The second sample consists of dark patches that contain Fermi sources with 
		an angular separation between 5$^\circ$ and 10$^\circ$ from the ROI center.
		}
\label{DarkPatches_flux}
\end{figure}

Using real data, downloaded from the Fermi Science Support 
Center \citep{ssc}, we again investigate the probability of false source detections due to diffuse background fluctuations, 
as in section \ref{falsedetections}.
The stacking is first performed with a sample of ROIs that contain as few Fermi-LAT detected sources as possible,
and next with ROIs that contain Fermi sources with 5$^\circ$ to 10$^\circ$ angular separation from the ROI center.
From all-sky data obtained during 162 weeks of LAT observations (2008-08-04 to 2011-09-13),
40 ROIs are selected for each of the two samples while galactic latitutes $|b|<25^\circ$ are again excluded. 
In the following, we denote these ROIs as dark patches. 
The same energy and ROI size cuts are applied as in section \ref{simulation} and the identical 
spacecraft information is used. 
The selected events belong to the \textit{SOURCE} 
class \citep{irf}. 
Before stacking, individual binned likelihood analyses are performed on each ROI, 
as it is described in section \ref{coadding}, and 
the Fermi-LAT detected point sources are simulated and subtracted from the data.
The resulting maps of counts (data minus simulated Fermi sources) are co-added.
For the co-adding analysis, a model is applied that includes the EGB, GB emissions and a power-law test source with photon index $-2.0$.
The normalizations of EGB and GB and the prefactor of the test source are, as in section \ref{simulation}, 
free parameters during the fitting procedure.
For both samples, the TS values computed for the test source remain $<1$ for all numbers of co-added ROIs.\\
\indent The resulting 90\% CL UL on the integrated test source flux are shown in figure \ref{DarkPatches_flux}.
A flux upper limit of approximately $3\times 10^{-11}$ ph/(cm$^2$s) is obtained after 40 stackings 
for the first sample, which is consistent with the results for the simulated 
dark regions in figure \ref{SimROI_flux}. In the second sample, the flux upper limit rises during the stacking and results in
approximately $5\times 10^{-11}$ ph/(cm$^2$s) after 40 co-additions. 
This behaviour can be explained by slight mis-modelings of the background point sources or 
the diffuse backgrounds in the vicinity of the source of interest.

\subsection{Detectability of weak signals}
\begin{figure}
\resizebox{\hsize}{!}{\includegraphics[width=\linewidth]{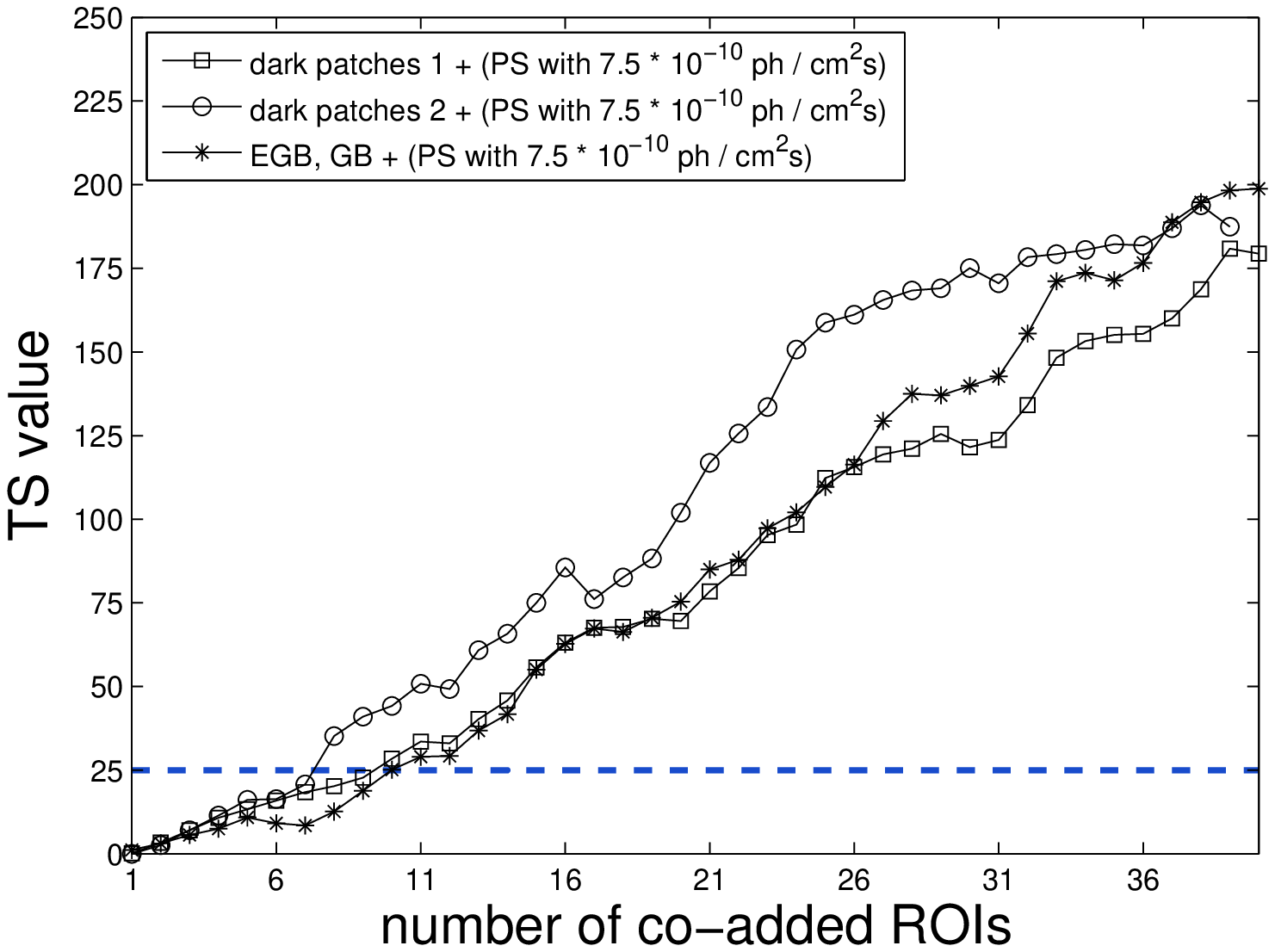}}
\caption{The test statistic values with respect to point-like emission at the center of the ROIs versus the number of co-added ROIs.
		The stacking is performed separately for three cases: simulated point-like sources (PS) with
		an integrated flux of $7.5\times 10^{-10}$ ph/(cm$^2$s) are added to 
		simulated diffuse background regions, the first sample of dark patches
		that contain as few sources as possible and to the second sample of dark patches
		that contain Fermi sources with 
		an angular separation between 5$^\circ$ and 10$^\circ$ from the ROI center.
		The detection threshold TS $\ge$ $25$ is indicated by the dashed line.}
\label{DarkPatches_CentralPS_TS}
\end{figure}

We perform a consistency check by repeating the study discussed in section \ref{weaksignals}, 
but this time the diffuse background is obtained from real data.
As before, a simulated point-like source with an integrated flux of $7.5\times 10^{-10}$ ph/(cm$^2$s) 
is added at the center of each dark patch. The detected Fermi sources are fit and subtracted from each ROI prior the stacking.
Using again a test source with photon index $-2.0$ in the model, we obtain a TS value for each co-adding step. The results are shown in 
figure \ref{DarkPatches_CentralPS_TS} and compared with the results previously obtained for the simulated diffuse backgrounds.
\begin{figure}
\resizebox{\hsize}{!}{\includegraphics[width=\linewidth]{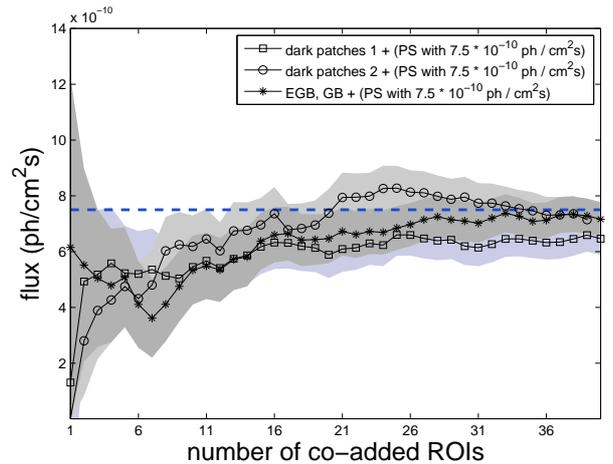}}
\caption{The integrated photon flux with respect to point-like emission at the center of the ROIs versus the number of co-added ROIs. 
		The stacking is performed separately for three cases: simulated point-like sources (PS) with
		an integrated flux of $7.5\times 10^{-10}$ ph/(cm$^2$s) (dashed line) are added to 
		simulated diffuse background regions, the first sample of dark patches
		that contain as few sources as possible and to the second sample of dark patches
		that contain Fermi sources with 
		an angular separation between 5$^\circ$ and 10$^\circ$ from the ROI center.
		}
\label{DarkPatches_CentralPS_flux}
\end{figure}
In all three cases, the detection threshold is reached within 5 to 10 stacking steps.\\
\indent The corresponding developments of the integrated flux values are shown in figure 
\ref{DarkPatches_CentralPS_flux}. The final values obtained after 40 co-additions are consistent with the input flux 
of $7.5\times 10^{-10}$ ph/(cm$^2$s).

\subsection{Application to real point-like emissions}
\label{realsources}
\begin{figure}
\resizebox{\hsize}{!}{\includegraphics[width=\linewidth]{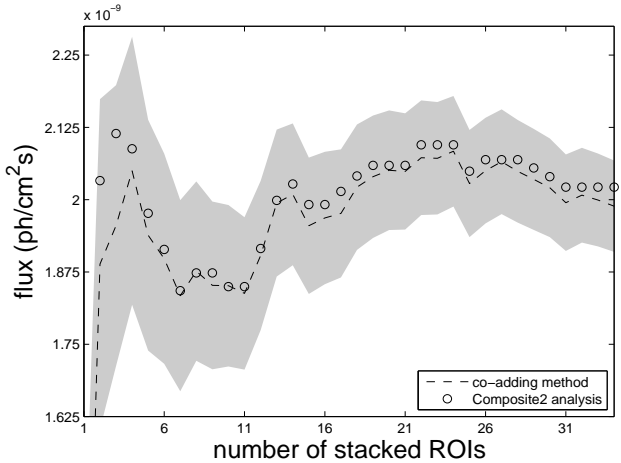}}
\caption{The integrated photon flux with respect to point-like emission at the center of the ROIs versus the number of co-added ROIs.
		The stacking is performed for Fermi-LAT detected point-like sources using the co-adding method 
		(the grey shaded area corresponds to the statistical uncertainty). The results obtained with the \textit{Composite2} analysis
		are plotted for comparison.
		}
\label{FermiSources_flux}
\end{figure}

In a further test, the method is applied to real point-like sources that are listed in the LAT 2-year point source catalog.
33 weak Fermi-LAT detected sources with average source significances close to 5$\sigma$ are selected for this purpose 
from galactic latitudes $|b|>25^\circ$.
As before, binned likelihood analyses are performed on the individual ROIs, in order to subtract the known
Fermi sources from the data. The 33 selected Fermi sources are treated as undetected and therefore not 
included in the models. 
For the co-adding analysis, a model is applied that includes EGB and GB emissions and
a point-like test source with photon index $-2.0$. We determine the TS values for each stacking step and find them 
approximately linearly increasing during the stacking, until a final value $\sim$1000 is reached after 33 co-additions. 
Figure \ref{FermiSources_flux} illustrates the development of the integrated test source flux during the stacking, 
which yields a final integrated flux of $2\times 10^{-9}$ ph/(cm$^2$s). 
We find an excellent agreement with the averaged integrated 
fluxes that are obtained from individual ROI analyses (no subtraction of simulated sources, no co-adding).\\
\indent As a further consistency check, we apply the stacking tool \textit{Composite2}, provided as part of the 
\textit{ScienceTools}, to the 33 Fermi sources. Since the ROIs are kept seperate in this case, 
we can not subtract 
the detected background point sources but need to provide individual models for each ROI that take into account 
these sources, 
the diffuse backgrounds as well as a test source at the center. The ROIs are then stepwise-added to the composite analysis 
and the integrated test source flux is determined after 
each stacking step. In figure \ref{FermiSources_flux}, the resulting integrated fluxes are compared to the values 
obtained with the co-adding method. The flux developments for both methods, the co-adding and the \textit{Composite2} analysis,
are in good agreement with each other, particularly for $>4$ stacking steps.
Due to the subtraction 
of point sources from the data, there is a departure between the co-adding and the \textit{Composite2} method 
during the first stacking steps, clearing away after a few co-additions since potential mis-modelings of the 
subtracted sources and resulting negative counts vanish in the diffuse background fluctuations.

\subsection{Visibility of stacked sources}
\begin{figure}
\resizebox{\hsize}{!}{\includegraphics[width=\linewidth]{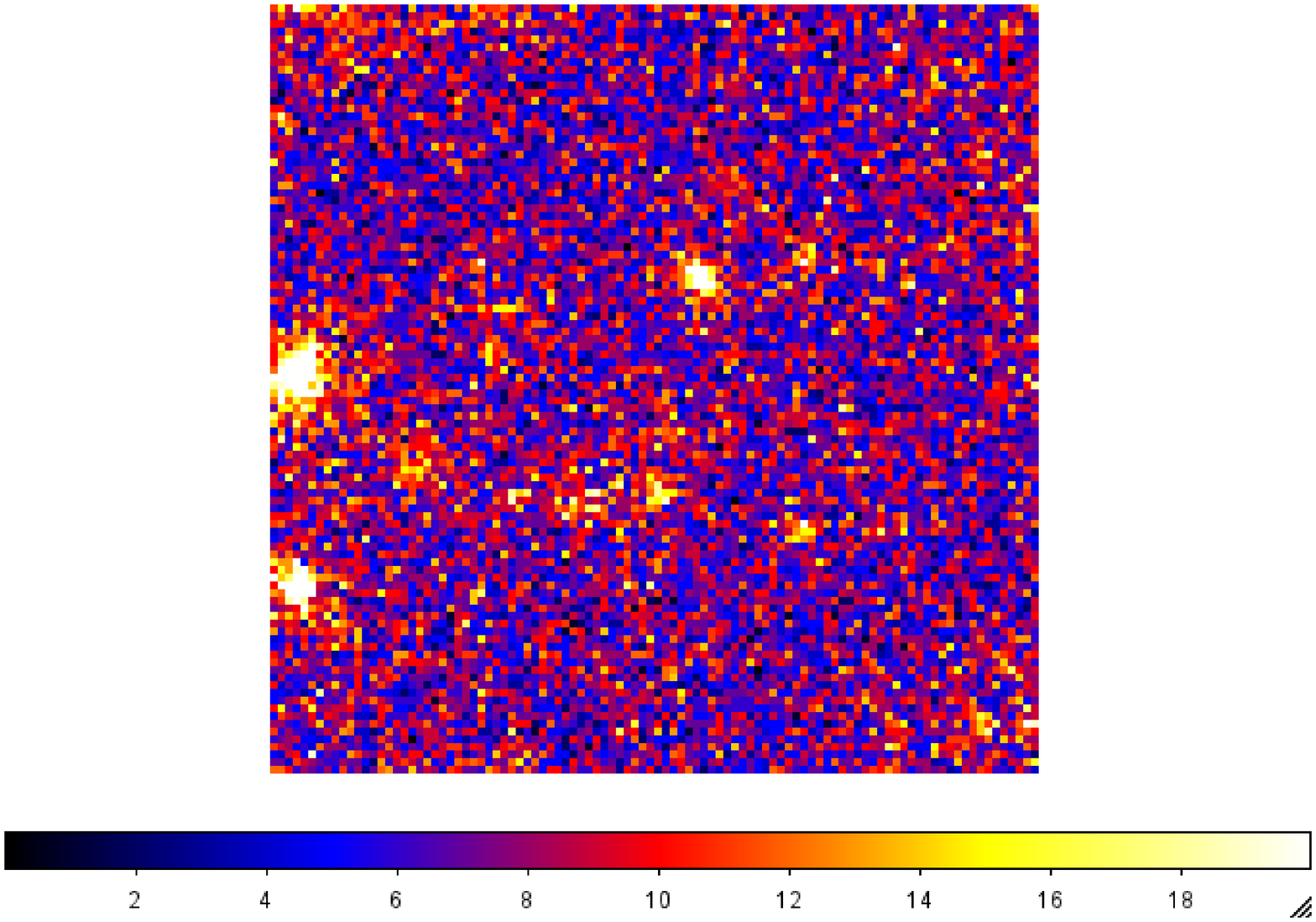}}
\caption{The map of gamma-ray counts of a region that hosts a weak but known point-like emission at its center.
		This map corresponds to the CountCube of this region, summed over all energies.
		}
\label{ds9_single3_1}
\end{figure}

\begin{figure}
\resizebox{\hsize}{!}{\includegraphics[width=\linewidth]{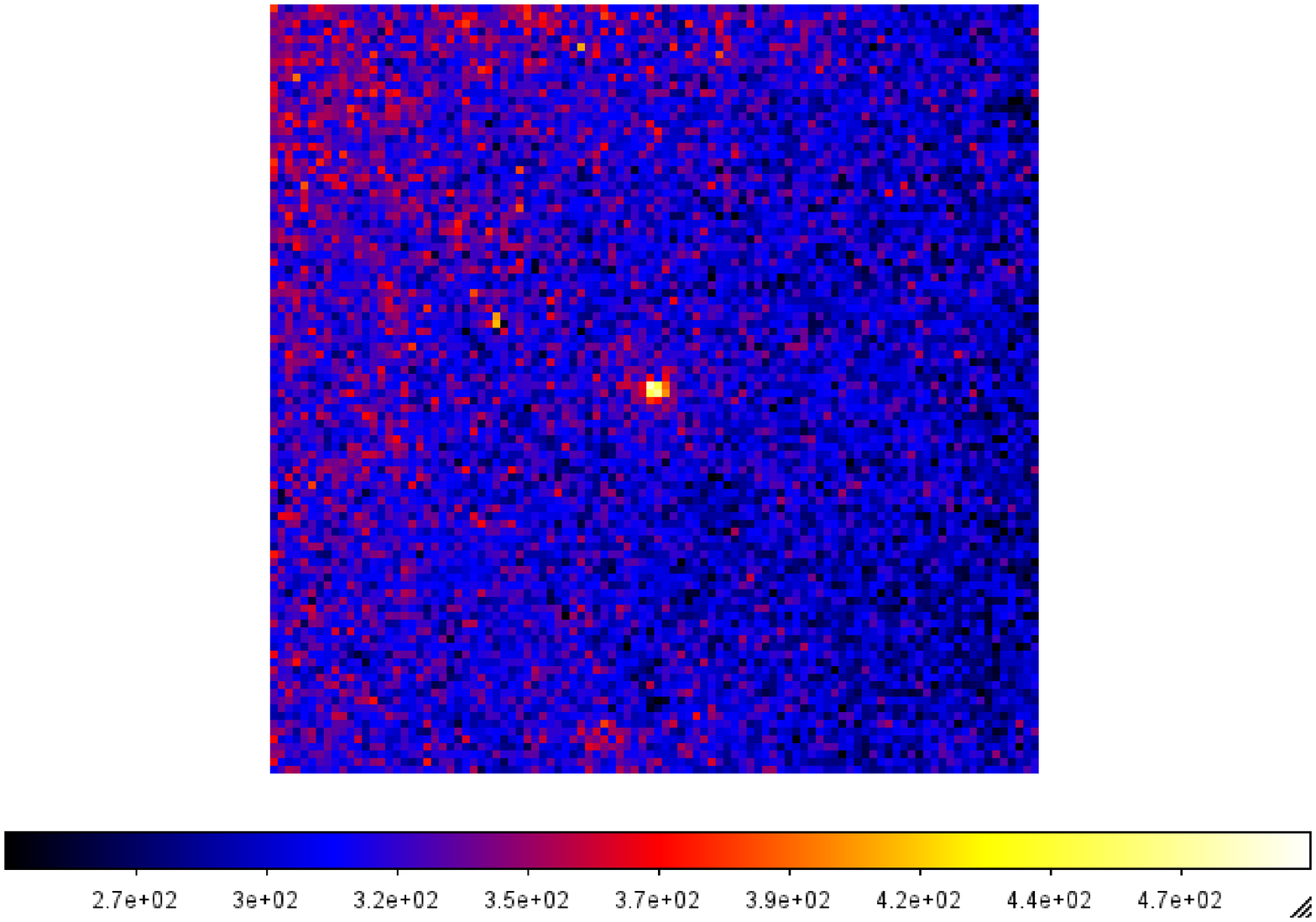}}
\caption{The map of gamma-ray counts of 33 stacked regions. Each of these regions hosts a weak but 
known point-like emission at the center.
		This map corresponds to the CountCube of these stacked regions, summed over all energies.
		}
\label{ds9_stacked3_33}
\end{figure}

The 33 Fermi-LAT sources from the previous section are used to illustrate the effect of the co-adding method on the maps of counts.
Figure \ref{ds9_single3_1} shows the map of counts for one of the 33 regions containing a weak but known point-like source at the center, 
as it is obtained from Fermi all-sky data. This map corresponds 
to the CountCube of this region summed over all energies. The simulated background 
point sources have not yet been subtracted. 
After preparing the 33 regions according to section \ref{coaddingdata}, i.e. by subtracting the non-central point sources, 
stacking of these regions results in the map of counts shown in figure \ref{ds9_stacked3_33}.
The background appears smooth, while the cumulative emission of the 33 sources is clearly visible at the center.

\section{Summary and discussion}
\label{discussion}
The stacking method presented in the previous sections is based on stacked maps of LAT counts and 
applies the public Fermi \textit{ScienceTools} for data preparation and maximum likelihood analysis. 
This method combines regions of potential gamma-ray sources in such a way that 
potential signals add constructively to increase the cumulative significance of these sources.
We demonstrate with the aid of simulations that the method is capable to detect weak point-like 
emissions from sources that are individually not significant and to determine the 
corresponding average photon flux and flux upper limits. The method is 
sensitive to the correct choice of the
spectral model for the sources to investigate, a feature that can be used for a systematic examination of 
the combined source spectrum.
We find that the stacking of hard emission spectra leads to a higher source significance compared to 
the stacking of soft spectra, due to the improvement of the Fermi LAT PSF at high energies. Furthermore, the source significance 
is generally higher if the individual spectral contributors are of similar spectral shape, and the
significance is decreased if the spectra deviate strongly from each other.
The method is successfully applied to real data, and an excellent agreement between the input and reconstructed 
source fluxes is found.\\
\indent Although the likelihood functions of the co-adding and the existing \textit{Composite2} method differ from each other, 
we show that both methods lead to similar results.
The co-adding of maps of counts allows background point sources to be subtracted from the data 
and the model-predicted contribution of the diffuse backgrounds to be combined. This leads to a simple model for the 
likelihood analysis, that
consists of only three components, i.e. the diffuse backgrounds and a common spectral model for the sources of interest, 
for any number of stacked sources. 
The co-adding method correlates the counts of the investigated sources in a defined way which can 
even help to make these sources visible in the count maps.

\onecolumn
\newpage
\bibliographystyle{aa}
\bibliography{references.bib}

\end{document}